\documentclass{aa}  

\usepackage{graphicx}
\usepackage{txfonts}
\usepackage{xcolor}
\begin{document} 

\title{Unbound grains in gaseous debris discs}

\subtitle{}

\author{Marija R. Jankovic \inst{1,}\thanks{Corresponding author: \email{marija.jankovic@ipb.ac.rs}},
      Tim D. Pearce \inst{2,3},
      Johan Olofsson \inst{4} \and
      Nicole Pawellek \inst{5} \and
      Mark C. Wyatt \inst{6}
      }

\institute{
    Institute of Physics Belgrade, University of Belgrade, Pregrevica 118, 11080 Belgrade, Serbia
    \and
    School of Physics, Trinity College Dublin, The University of Dublin, College Green, Dublin 2, Ireland
    \and
    Department of Physics, University of Warwick, Gibbet Hill Road, Coventry CV4 7AL, United Kingdom
    \and
    European Southern Observatory, Karl-Schwarzschild-Strasse 2, 85748 Garching bei München, Germany
    \and
    Institut für Astrophysik, Universität Wien, Türkenschanzstraße 17, 1180 Wien, Austria
    \and
    Institute of Astronomy, University of Cambridge, Madingley Road, Cambridge CB3 0HA, United Kingdom
         }

\date{Received 29 May 2026; accepted 15 September 2026}

 
\abstract
{Some debris discs contain cold molecular gas. The origin of this gas is one of the key questions in our understanding of the evolution of planetary systems. The effect of gas on the smallest dust grains, which are expected to be unbound due to radiation pressure, is not well understood. The unbound grains contribute to the brightness of scattered light and thermal emission from the disc halo at short wavelengths.}
{We explored the conditions under which gas can prevent or slow down the escape of unbound grains and increase their number density in the disc halo sufficiently to affect its brightness and reveal properties of the gas.}
{We numerically integrated the equations of motion of dust grains released on unbound orbits in optically thin, gaseous debris discs, taking into account stellar gravity, radiation pressure, Poynting-Robertson drag, and gas drag.}
{We find that gas drag can significantly slow down the escape of unbound grains only for very high gas masses, such as those expected if the gas is remnant protoplanetary gas and not of secondary origin. We also present an analytical model for estimating the density of unbound grains in gaseous discs.}
{If a difference in the abundance of unbound grains between gaseous and non-gaseous debris discs were observed, it would imply that the gas in debris discs is protoplanetary. Currently, there is no evidence of such a difference, suggesting that the gas detected in debris discs is either secondary or strongly depleted protoplanetary gas.}

\keywords{minor planets, asteroids: general -- circumstellar matter -- protoplanetary discs}

\maketitle
%

\section{Introduction}
Minor bodies in planetary systems are largely concentrated in belts or discs and are most commonly called debris discs \citep{Wyatt2008, Krivov2010, Matthews2014, Hughes2018, Wyatt2020}. Debris discs around stars other than the Sun are studied by observing their optically thin dust, which is believed to be produced in collisions of the larger bodies, the planetesimals. Modern observational facilities allow us to study thermal emission and scattered stellar light from sub-micron to centimetre-sized dust grains in great detail. For example, the Spectro-Polarimetric High contrast imager for Exoplanets REsearch (SPHERE) and the Atacama Large Millimeter/submillimeter Array (ALMA) provide such observations \citep{Engler2025,Marino2026a}. The planetesimals producing the dust in debris discs are thought to be by-products of planet formation occurring during an earlier, protoplanetary phase of the system. In the protoplanetary phase, the star is surrounded by an optically thick disc of primordial dust and gas dominated by molecular hydrogen, and planetesimals and planets may or may not have formed yet \citep{Beckwith1990,Williams2011}. In contrast, the second-generation dust in debris discs is optically thin, and debris discs are typically gas-poor. The transition from the protoplanetary to the debris disc phase involves the removal of primordial dust and gas and the onset of planetesimal collisional evolution, but the system age and the order in which these processes occur is not fully understood \citep{Wyatt2015}.

Some systems whose dusty component classifies them as debris discs are immersed in a cold gaseous disc, detected mainly through thermal emission from molecular CO, with CO frequently being as abundant as in some protoplanetary discs \citep[e.g.][]{Kospal2013, Dent2014, Lieman-Sifry2016, Moor2017, Higuchi2017, Cataldi2018,Kral2020}. One of the key questions in current debris disc research is whether the cold gas is second-generation gas released from the planetesimals \citep{Zuckerman2012,Dent2014,Bonsor2023}, or whether the CO is merely a tracer of hydrogen-rich gas that is still present from the protoplanetary phase \citep{Kospal2013, Moor2017}. The answer to this question will provide information on the internal composition of planetesimals, their collision rates,  surface processes, or protoplanetary disc dispersal processes.

Due to a combination of observational selection effects, debris discs are most easily detected around A-type stars \citep[e.g.][]{Thureau2014,Sibthorpe2018,Lestrade2025}. Around A-type stars, sub-micron to micron-sized dust is expected to be produced on unbound orbits and to escape quickly due to stellar radiation pressure, making these small dust grains much less numerous than the larger, bound grains \citep[e.g.][]{Krivov2006}. Theoretical studies on the effects of gas drag in debris discs have therefore mostly focused on studying the much more abundant bound dust grains. Bound dust grains in gaseous discs may have their orbital inclinations and eccentricities damped, and they may migrate radially, depending on the grain size and gas density. They may settle on so-called stationary orbits, stellocentric radii at which radiation pressure, Poynting-Robertson (PR) drag, and gas drag forces are in equilibrium \citep{Takeuchi2001,Pearce2020}. These effects act on timescales shorter than the collisional lifetimes of dust grains for a wide range of gas densities. They may shape the halo of gaseous debris discs \citep{Thebault2005,Krivov2009}, alter the vertical structure of the disc \citep{Olofsson2022}, and produce additional dusty rings \citep{Takeuchi2001,Jankovic2026} and asymmetric structures \citep{Weber2026}. The Two-way interaction between the gas and the dust, which accounts for the back-reaction on the gas, can reshape the structure of the disc to an even greater degree \citep{Klahr2005,Besla2007,Lyra2013,Castrejon2019}.

Although bound grains contribute the most to the brightness of debris discs, the contribution of unbound grains is not negligible \citep{Thebault2019,Thebault2023}. Additionally, for gaseous discs, several studies have hypothesised that gas drag can slow unbound grains or even cause them to become bound in the system \citep[e.g.][]{Bhowmik2019, Moor2019, Milli2023}. However, so far there have been few theoretical studies exploring the fate of unbound grains in the presence of gas. \citet{Takeuchi2001} briefly noted that unbound grains are only affected by the gas if the drag stopping time is shorter than the dynamical timescale. Assuming that the escape of unbound grains can be fully halted under such a condition, \citet{Pawellek2024} find a good model fit to the morphology of the disc around HD~131488, an A-type star. However, it is not clear whether the escape can truly be prevented. \citet{Pearce2020} included unbound grains in their study of close-in hot dust and gas produced from dust sublimation and found that the escape of unbound dust grains is not halted for the gas densities considered therein, although the escape is slowed down. They also found that the grains continue to move more slowly even after they leave the close-in gaseous disc. For cold, CO-rich gaseous debris discs, \citet{Olofsson2026} found that unbound grains can contribute non-negligibly to disc brightness in the near-IR and mid-IR when gas is present. However, they do not discuss the dynamics of unbound grains and the dependence on model parameters in detail. The only study dedicated to dynamics of unbound grains under the influence of gas drag appears to be \citet{LecavelierDesEtangs1998}, who considered an outflow of unbound particles from a debris disc into a gaseous interstellar environment. \citet{LecavelierDesEtangs1998} explored in detail the three phases of such an outflow: (1) acceleration due to radiation pressure, (2) deceleration after gas drag kicks in, and (3) stationary outflow after radiation pressure and gas drag balance out. However, these results are not directly applicable to gaseous debris discs that are significantly more compact, in which the gas density varies by many orders of magnitude as a function of distance to the star, and in which the observationally most significant (i.e. the brightest) unbound particles are those with sizes just below the bound limit.

In this paper we show that gas drag can significantly enhance the surface brightness of unbound grains in debris discs only in very gas-rich discs, which offers insights into the origin of gas in debris discs. We also present an approximate analytical approach to estimating the density of unbound grains. In Sections \ref{sec:methods}, \ref{sec:results}, and \ref{sec:conclusions} we present our methods, results, and conclusions, respectively.

\section{Methods} \label{sec:methods}
We considered a narrow belt of planetesimals on circular orbits at a distance of $r_{\rm belt}=100$\,au from a star. We assumed the star to be an A-type star with a mass $M_\star=1.7$\,M$_\odot$, luminosity $L_\star=8.5$\,L$_\odot$, radius $R_\star=1.5$\,R$_\odot$, and a black-body spectrum corresponding to an effective temperature of $T_\star=7951$\,K. These are the properties of the archetypal debris-disc host $\beta$~Pic \citep[taken from][]{Marino2026a}, but we did not aim to reproduce the properties of the disc around $\beta$~Pic specifically. We assumed that planetesimals feed a collisional cascade that produces dust grains sufficiently small to be strongly affected by radiation pressure and to be released on unbound orbits. We assumed that dust grains are released from circular orbits at $r_{\rm belt}$ and are therefore unbound if their $\beta$ parameter, the ratio of radiation pressure to stellar gravity force, satisfies $\beta > 0.5$.

\begin{figure}
    \centering
    \includegraphics[width=\columnwidth,clip,trim={0cm 3cm 0cm 3cm}]{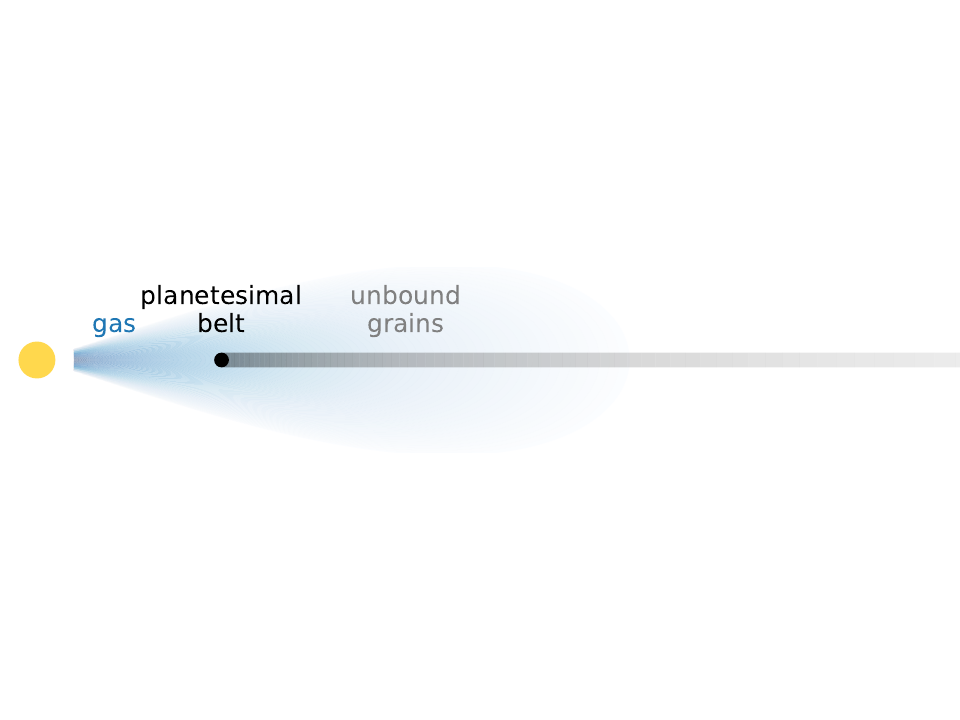}
    \caption{Sketch of the system, showing the star (yellow circle), gas disc (blue), and planetesimal belt (black circle) in which the unbound grains (grey) are produced. The gas may or may not extend much further inwards from the planetesimal belt; this is irrelevant for the unbound grains, which only move outwards.}
    \label{fig:sketch}
\end{figure}

We further assumed that the dusty disc is immersed in a gaseous disc, as depicted in Fig. \ref{fig:sketch}. As discussed above, the gas in debris discs may be either second-generation gas released from planetesimals or remnant protoplanetary gas. In either case, we assumed that the gas surface density is given by the self-similar solution \citep{Lynden-Bell1974}, $\Sigma(r) \propto ( r/r_{\rm belt} )^{-1} e^{-r/r_{\rm belt}}$, normalised such that the integrated mass from $r_{\rm belt}$ to infinity equals a gas mass $M_{\rm g}$. We assumed that the gas also extends inwards of $r_{\rm belt}$, but this is inconsequential for the unbound grains, which only move outwards. We further assumed a black-body radial temperature profile for the gas, a vertically constant temperature, and a vertical hydrostatic equilibrium. The midplane gas density is then given by $\rho(r)=\Sigma(r)/(\sqrt{2\pi} H)$, where $H=c_{\rm s}/\Omega_{\rm K}$ is the gas scale height, $c_{\rm s}$ is the isothermal sound speed, and $\Omega_{\rm K}$ is the Keplerian angular velocity at the given radius \citep[e.g.][]{Armitage2010}. 

To model the two possible origins of the gas, we varied the gas mass, $M_{\rm g}$, and the mean molecular weight, $\mu_{\rm g}$. If the cold CO detected in debris discs is second-generation gas, the gas mass and its composition are expected to be dominated by the observed CO, or possibly by its photodissociation products, carbon and oxygen. Observations show up to $\sim 0.1$\,M$_\oplus$ of CO in debris discs \citep[from the data compiled in the appendix of][]{Marino2020}. The mass of carbon exceeds the mass of CO in some systems, but never reaches $0.1$\,M$_\oplus$ (based on the same dataset). We therefore adopted this mass as the upper limit for the gas mass in the second-generation scenario. We took the mean molecular weight of the gas in this scenario to be that of CO, $\mu_{\rm g}=28$, in units of the hydrogen atom mass. Alternatively, if the detected CO is only a tracer of protoplanetary amounts of molecular hydrogen, the actual gas mass is expected to be significantly higher. Assuming an H$_2$-CO number density ratio of $\sim 10^4$ (typical of protoplanetary discs), the total gas mass in gaseous debris discs is approximately $10^3$ times higher than the mass of the detected CO, which places the upper limit of the total gas mass in debris discs at $\sim 100$\,M$_\oplus$. In general, masses of protoplanetary discs are estimated to range up to a few percent of the stellar mass \citep[or a few thousand M$_\oplus$;][]{Williams2011}. For protoplanetary gas, we set the mean molecular weight to $\mu_{\rm g}=2.3$.

We followed the dynamics of the unbound grains in the optically thin gaseous disc, taking into account stellar gravity, radiation pressure, PR drag, and gas drag. For simplicity, we assumed that the dusty disc is vertically thin, so that particles move in the gas disc midplane. We integrated the equations of motion given by \citet{Takeuchi2001} using the \texttt{solve\_ivp} method from the \texttt{python} library \texttt{scipy}, which implements the Runge-Kutta 5(4) integrator with an adaptive time step \citep[fifth order integrator with fourth order integrator used for the error estimate;][]{Dormand1980}. We set the maximum time step to $2\pi/(100 \beta \Omega_{\rm K}(r_{\rm belt}))$, where $\Omega_{\rm K}(r_{\rm belt})$ is the Keplerian angular velocity at $r=r_{\rm belt}$. We typically integrated for either $10^6$\,yr or until a particle reached $10^5$\,au, whichever occurred sooner. Where relevant, we additionally checked that a particle had travelled beyond the region of interest and integrated for longer if not (this occurs when particles are extremely well coupled to the gas), and then we checked that the particle is indeed gravitationally unbound and will not return towards the star.

We used \texttt{optool} \citep{optool} to calculate the optical properties of the dust. We applied Mie theory and assumed the grain composition to be that of compact (nonporous) astrosilicate \citep[a mixture of amorphous rocky material developed as a model of dust grains in the interstellar medium;][]{Draine2003} and we adopted a dust-grain bulk density of $\rho_{\rm s}=3.3$\,g\,cm$^{-3}$.  We made these choices for simplicity following previous theoretical work \citep[e.g. the study of unbound grains by][]{Thebault2019}. With these assumptions and the adopted stellar properties, the blowout limit (the size for which $\beta=0.5$) is $s_{\rm blowout}\approx 2$\,\textmu m. Overall, the $\beta$ parameter has a non-monotonic dependence on particle size, peaking around 6.5 at a size of $\sim 0.1$\,\textmu m \citep[as illustrated by Fig. 1 in][but we assumed a slightly higher stellar luminosity, which yields slightly higher $\beta$ values at a given grain size and a slightly higher blowout limit]{Thebault2019}. The non-monotonic dependence arises because dust grains smaller than the wavelength of peak stellar emission absorb and scatter stellar light poorly, which reduces their radiation pressure efficiency and their $\beta$ parameter.

Assuming that unbound particles of size $s$ are produced at a rate $\dot{n}(s)$ per unit size and leave the system with a radial velocity $v_r(s,r)$, the number of particles per unit area (the vertical number column density) per unit size is given by \citep[e.g.][]{vanLieshout2014}
\begin{equation} \label{eq:num_dens}
    n(s,r) = \frac{1}{2\pi r} \frac{\dot{n}(s)}{v_r(s,r)}.
\end{equation}
The production rate is determined by the properties of the collisional cascade \citep[e.g. the total debris disc mass, the size of the largest planetesimals, the disc excitation, etc.,][]{Wyatt2007b,Thebault2019}. In this work we are only interested in the relative changes to the number density of unbound grains compared to a disc with no gas, and so we did not calculate the exact value of the production rate. For our calculations, the relevant relation is the dependence of $\dot{n}$ on particle size, which we assume follows a power-law, $\dot{n}(s) \propto s^{-q}$, with $q=3.5$. The radial velocity $v_r(s,r)$ is the key quantity extracted from numerical integration.

The contribution of unbound grains to the surface brightness of scattered light is then given by
\begin{equation} \label{eq:sb_unbound}
    I_{\rm sca, unbound}(r) \propto \int_{s_{\rm min}}^{s_{\rm max}} \frac{1}{r^3} s^{2-q} Q_{\rm sca}(s,\lambda) F_\star(\lambda) \frac{1}{v_r(s,r)} \textrm{d}s,
\end{equation}
where we set $s_{\rm min}=0.01$\,\textmu m and $s_{\rm max}$ as the size above which particles become bound, which is model-dependent, but always very close to the blowout size limit, $s_{\rm blowout} \approx 2$\,\textmu m. Furthermore, $F_\star(\lambda)$ is the stellar flux and $Q_{\rm sca}(s,\lambda)$ is the scattering coefficient. We assumed isotropic scattering to isolate dynamical effects from those due to dust optical properties \citep[this is a good approximation for very small dust grains, see Fig. A2 of][]{Pawellek2024}. We used a grid of 100 sizes between $s_{\rm min}$ and the blowout size. Analogously, the contribution of unbound grains to the surface brightness of thermal emission is given by
\begin{equation} \label{eq:sb_unbound_therm}
    I_{\rm therm, unbound}(r) \propto \int_{s_{\rm min}}^{s_{\rm max}} \frac{1}{r} s^{2-q} Q_{\rm abs}(s,\lambda) B(T(s,r),\lambda) \frac{1}{v_r(s,r)} \textrm{d}s,
\end{equation}
where $Q_{\rm abs}(s,\lambda)$ is the absorption coefficient and $B(T,\lambda)$ is the Planck function. The temperature $T$ is a function of grain size and radius and is calculated from thermal equilibrium, assuming the stellar spectrum is well approximated by the Planck function at the adopted $T_\star$.

\section{Results} \label{sec:results}
We considered dust grains with sizes ranging from $s=0.01$\,\textmu m ($\beta = 0.8$) to the blowout size $s \approx 2$\,\textmu m ($\beta=0.5$; note that, as discussed above, $\beta$ peaks around $6.5$ at a size of 0.1\,\textmu m for the adopted grain composition and stellar parameters). The grains are embedded in either CO-dominated gas (with a mean molecular weight of $\mu_{\rm g}=28$, in units of the hydrogen atom mass), or hydrogen-dominated gas ($\mu_{\rm g}=2.3$), with a gas mass of up to 100\,M$_\oplus$ in the region outwards of the radius $r_{\rm belt}=100$\,au, at which we released the dust grains. We find that the vast majority of dust grains produced on initially unbound orbits remain unbound and escape the system, although they take longer to do so than in the absence of gas. We discuss the decrease in the radial velocity due to gas drag in Sect. \ref{sec:velocity} and the condition for particles to become bound in Sect. \ref{sec:trapping}. In Sect. \ref{sec:surface_brightness} we present the results on the effect of gas on the surface brightness of unbound grains.

\subsection{Radial velocity evolution of unbound grains} \label{sec:velocity}
\begin{figure*}
    \centering
    \includegraphics[width=\textwidth]{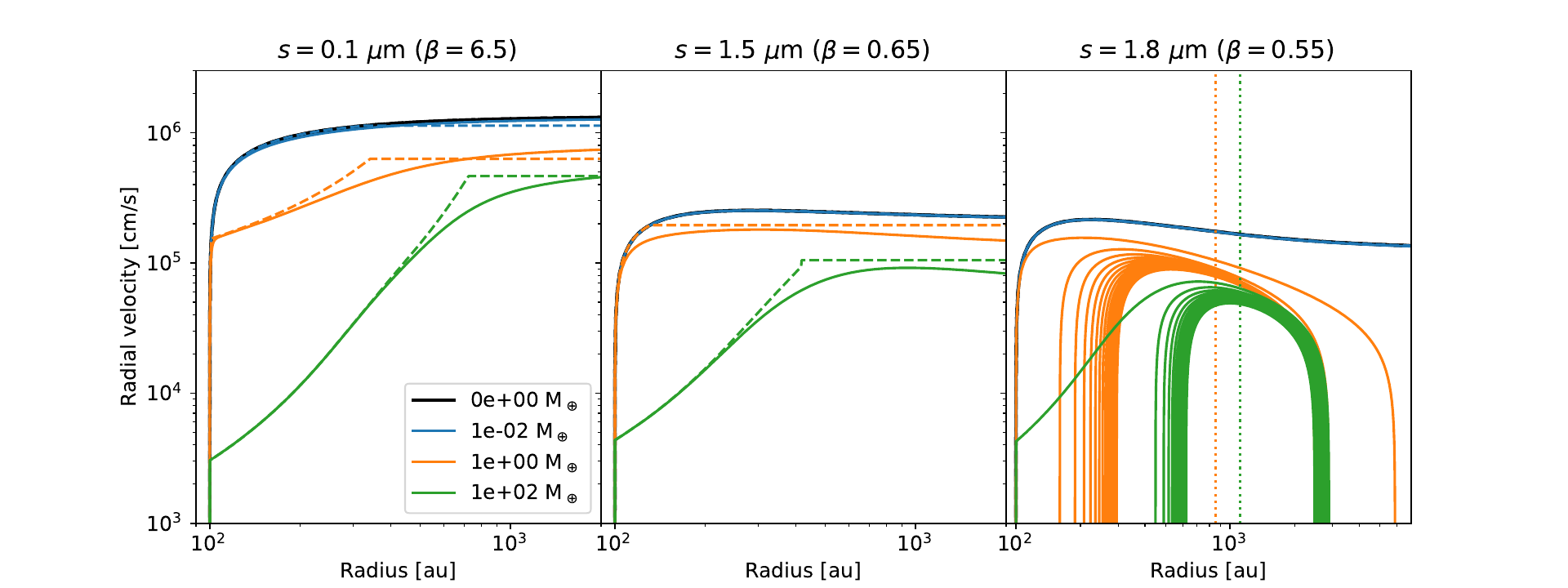}
    \caption{Radial velocity as a function of radius for dust grains of three different sizes (see panel titles) in a disc with no gas and in gaseous discs with three different gas masses (different colours; see plot legend). Solid lines show the results of numerical integration of the equations of motion, while dashed lines show the analytical model. The particles with $\beta=0.55$ become bound if the gas is sufficiently massive. In these cases (orange and green curves in the right-hand panel), the curves wrap around multiple times as the particles repeatedly oscillate radially inwards and outwards along eccentric orbits. Periods during which the particles move inwards (negative radial velocity) are not shown because of the logarithmic axis. The curves wrap around progressively more tightly as the orbits slowly circularize. We show the equilibrium radii at which these particles would settle given sufficient time as vertical dotted lines. The gas is assumed to be composed primarily of molecular hydrogen.}
    \label{fig:velocity}
\end{figure*}

In Fig. \ref{fig:velocity} we show the radial velocity obtained by integrating the equations of motion for dust grain sizes $s=0.1$--$1.8$\,\textmu m ($\beta=6.5$--$0.55$) and for various gas masses. The initially marginally unbound dust grain with $s=1.8$\,\textmu m ($\beta=0.55$, right-hand panel) becomes bound for a sufficiently large gas mass. We discuss the conditions for this to occur in the next section. In this section we focus on the radial velocity of the smaller dust grains. The radial velocity of these particles is reduced by the gas at all radii, yet they remain unbound and continue moving outwards. Here we assumed a hydrogen-dominated gas, but the same conclusions apply for a CO-dominated gas, and these results are also consistent with studies of hot dust in gas produced in the innermost regions of planetary systems, at the dust sublimation line \citep{Pearce2020}.

The dashed curves in Fig. \ref{fig:velocity} show an analytical estimate of the radial velocity at which particles escape. In the absence of gas or in the low-gas-mass limit, the radial velocity of particles is well described by \citep[e.g.][]{vanLieshout2014}
\begin{equation} \label{eq:v_r_nogas}
    v_{\textrm{no gas}}(r) = \sqrt{ \frac{G M_\star}{r} \left( 2(1-\beta) - \frac{r}{r_{\rm belt}} (1-2\beta) - \frac{r_{\rm belt}}{r} \right) } .
\end{equation}
The dust grains are initially rapidly accelerated radially, after which their radial velocity asymptotes towards
\begin{equation} \label{eq:v_r_nogas_inf}
    v_{\textrm{no gas}, \infty} = v_{\rm K}(r_{\rm belt}) \sqrt{2 \beta - 1},
\end{equation}
where $v_{\rm K}$ is the Keplerian velocity.

In the presence of gas, the velocity is well described by an analytical model composed of three components. We first consider motion through the gaseous disc. A significant decrease in the radial velocity occurs only if the gas density is sufficiently high for the particle stopping time to be less than the dynamical timescale, that is, if the dimensionless stopping time is $T_{\rm s} < 1$ \citep{Takeuchi2001}. Particles couple to the gas azimuthally, so that their azimuthal velocity is dragged towards that of the gas rotation, $v_{\rm g}=v_{\rm K}\sqrt{1-\eta}$, where $\eta$ is the ratio of the pressure gradient force to the gravitational force acting on the gas. In the radial direction, particles accelerate to a terminal velocity \citep{Takeuchi2001},
\begin{equation} \label{eq:v_r_with_gas}
    v_{\rm with\,gas}(r) \approx (\beta - \eta) T_{\rm s} v_{\rm K}.
\end{equation}
Escaping particles reach supersonic velocities, so the dimensionless stopping time $T_{\rm s}$ depends on the particle velocity. The solution for the velocity is \citep[for $T_{\rm s} < 1$,][their equation 18]{Takeuchi2001}
\begin{align} \label{eq:v_r_term}
    v_{\rm with\,gas} = &\textrm{sgn}(\beta - \eta) \frac{v_{\rm T}}{\sqrt{2}} \nonumber \\ 
    & \times \left\{ \left[ 1 + \left( 2 T_{\rm ss} (\beta - \eta) \frac{v_{\rm K}}{v_{\rm T}} \right)^{2} \right]^{1/2} - 1 \right\}^{1/2},
\end{align}
where $v_{\rm T}$ is the mean thermal velocity, $T_{\rm ss}$ is the dimensionless stopping time in the subsonic limit, and sgn denotes the signum function. Because $\beta > \eta$ for the small dust grains, particles move outwards in the presence of gas. As the dust grains move radially outwards, the terminal velocity increases because the gas density drops and $T_{\rm s}$ increases.

At large distances where $T_{\rm s} > 1$, the terminal velocity is expected to decrease if particles are bound \citep{Takeuchi2001}. The unbound particles instead become decoupled from the gas and continue to escape as a result of radiation pressure. A particle's velocity at infinity is reduced compared to that in a gas-less disc due to energy loss caused by gas drag, as previously shown by \citet{Pearce2020}. We find that a reasonably good approximation for this velocity is obtained by modifying Eq. (\ref{eq:v_r_nogas_inf}),
\begin{equation} \label{eq:v_r_with_gas_infty}
    v_{\textrm{with\,gas}, \infty} = v_{\rm K}(r(T_{\rm s} = 1)) \sqrt{2 \beta - 1},
\end{equation}
which is the velocity at infinity under the influence of radiation pressure for a particle released from a circular orbit at the radius where $T_{\rm s}=1$. Overall, the radial velocity of dust particles is well described by the minimum of Eqs. (\ref{eq:v_r_nogas}), (\ref{eq:v_r_term}), and (\ref{eq:v_r_with_gas_infty}) at any given radius, accounting for the initial acceleration phase, steady-state drift due to gas drag, and reduced velocity at large distances. The analytical result deviates from the numerical result only at radii where particles start to decouple from the gas, where $T_{\rm s}$ exceeds unity and Eq. (\ref{eq:v_r_term}) becomes invalid.

\subsection{Trapping of marginally unbound grains} \label{sec:trapping}
Some particles that are initially produced on marginally unbound orbits can become bound and remain in the system (see the right-hand panel of Fig. \ref{fig:velocity}), but the `trapping' by gas is very limited. Taking into account radiation pressure, the energy per unit mass of a dust grain is given by
\begin{equation}
    \epsilon = - \frac{G M_\star (1-\beta)}{r} + \frac12 v^2.
\end{equation}
If a particle is released from a circular orbit at radius $r$ with azimuthal velocity $v_{\rm K}(r)$, then, using the virial theorem, $\epsilon = - v_{\rm K}(r) (1/2 - \beta)$. This is a well-known result: particles released from planetesimals on circular orbits escape if $\beta > 1/2$ \citep[e.g.][]{Krivov2006}. Particles with $\beta > 1$ can never be trapped by gas drag, because even at zero kinetic energy, the radiation pressure is stronger than gravity and particles still escape. Particles with $1/2 < \beta < 1$ can become bound if the gas drag sufficiently reduces their kinetic energy. 

However, we find that few particles with $1/2 < \beta < 1$ become bound. We numerically integrated the equations of motion for particles in the corresponding size range, assuming a gas mass of 100\,M$_\oplus$ (defined as the mass in the region outwards from the belt radius, as above).
We calculated the specific energy of particles at the end of the simulation and find that only particles with $\beta \lesssim 0.6$ become bound for the gas quantities tested. For a gas mass of 1000\,M$_\oplus$ (ten times higher), only particles with $\beta \lesssim 0.65$ become bound. The difference in the corresponding `blowout' size in the presence of gas compared to a gas-free disc is extremely small relative to various uncertainties in dust optical properties and is virtually undetectable.

The evolution of $\beta > 0.5$ particles that are bound due to gas drag is similar to that of particles with $\beta < 0.5$. These particles drift radially at the terminal velocity given by Eq. (\ref{eq:v_r_with_gas}), towards a radius at which the radiation pressure, gas drag, and PR drag forces balance\citep{Takeuchi2001,Pearce2020,Jankovic2026}. Particles with high $\beta$ are expected to reach the outer regions where the gas density is reduced, where they may accumulate to form a secondary dusty ring \citep[][]{Takeuchi2001}. However, at the outer edge of the gaseous disc, the orbits of the particles also become highly eccentric, circularising very slowly over time \citep{Takeuchi2001}. This behaviour is seen in the right-hand panel of Fig. \ref{fig:velocity} for $\beta=0.55$ and a gas mass of 1\,M$_\oplus$ and 100\,M$_\oplus$ (orange and green curves). The equilibrium radii in these two cases are shown by the vertical dashed lines. They depend on where the gas density drops sufficiently for the forces to balance, and in a higher-gas-mass disc the equilibrium radius is further out. In each case, the particle initially moves outwards with increasing radial velocity. At some point the radial velocity starts decreasing and, after the particle travels beyond the equilibrium radius, its radial velocity reaches zero at the orbital apocentre. The particle then moves back towards the star. This inward track, where the radial velocities are negative, is not shown in Fig. \ref{fig:velocity} because of its logarithmic (and therefore positive) y axis, but the track eventually reverses at the orbit's pericentre and reappears in the figure. The curves wind tighter and tighter around the equilibrium radius, which depicts slow orbit circularisation.

\subsection{Contribution to disc brightness} \label{sec:surface_brightness}

\begin{figure*}[h!]
    \centering
    \includegraphics[width=\linewidth]{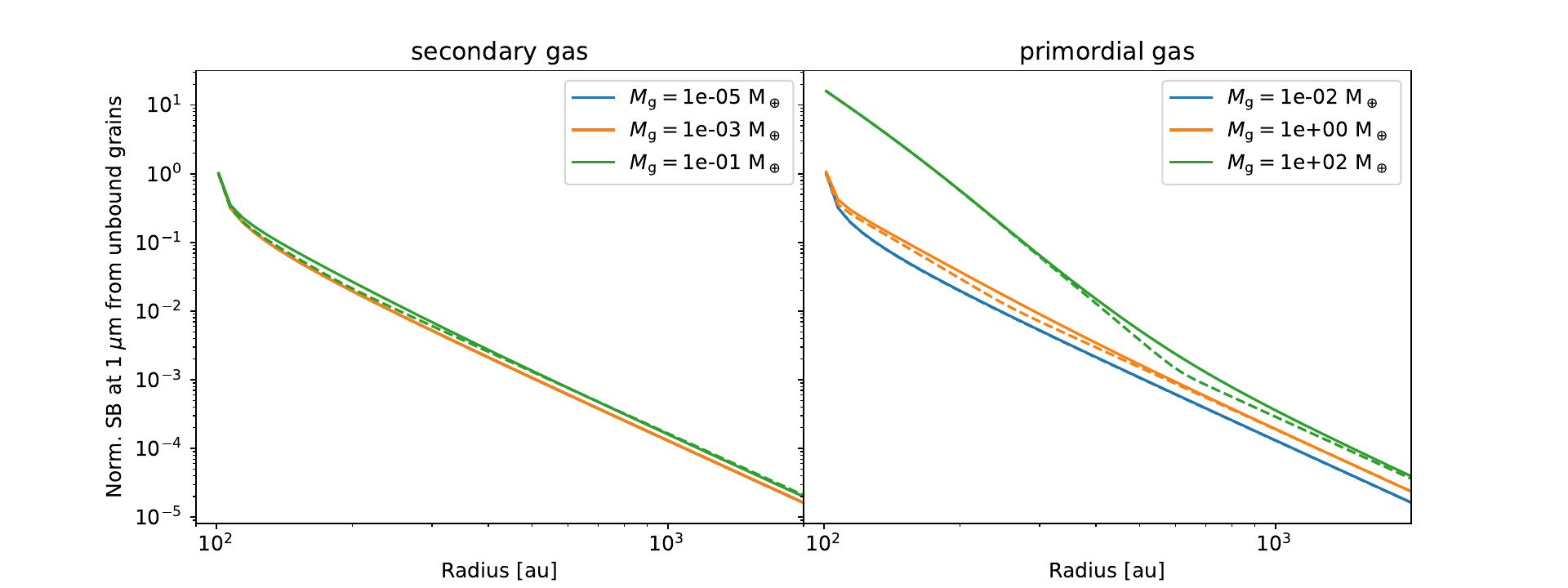}
    \caption{Contribution of unbound grains to the disc surface brightness in scattered light at 1\,\textmu m as function of radius, from the numerical model (solid lines) and an analytical model (dashed lines). The values are normalised so that the surface brightness at 100\,au is equal to one in the low-gas-mass limit. Left: Gas is assumed to be composed primarily of CO. Right: Gas is assumed to be composed primarily of molecular hydrogen and to have a much higher mass.}
    \label{fig:sb_1um}
\end{figure*}

\begin{figure*}[h!]
    \centering
    \includegraphics[width=\linewidth]{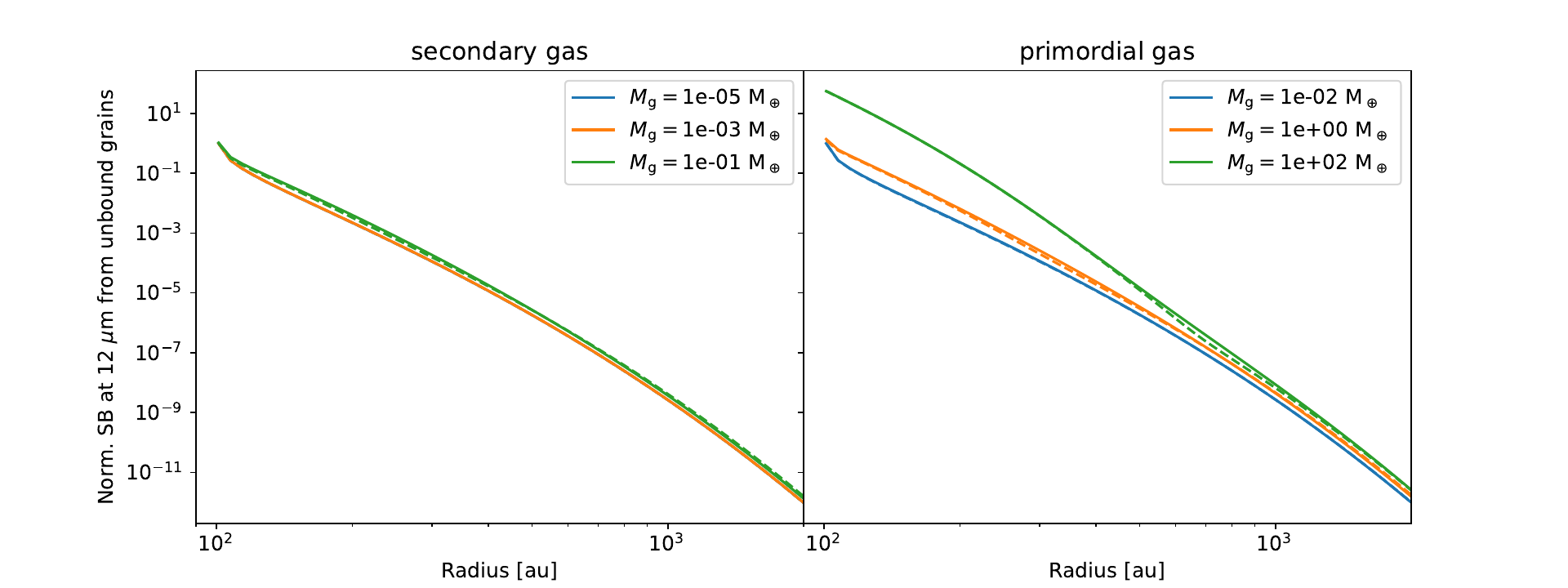}
    \caption{Same as Fig. \ref{fig:sb_1um} but for thermal emission at 12\,\textmu m.}
    \label{fig:sb_12um}
\end{figure*}

We calculated the contribution of unbound grains to the surface brightness of scattered light at a wavelength of 1\,\textmu m using Eq. (\ref{eq:sb_unbound}) and to the surface brightness of thermal emission at 12\,\textmu m using Eq. (\ref{eq:sb_unbound_therm}), for various gas parameters. We did not include particles that become bound and trapped at the outer gas disc edge ($\beta \lesssim 0.6$ for our most massive gas disc; see the previous section). This is because the focus of this work is on the dynamics of unbound grains, even if bound particles are naturally expected to contribute much more to the total surface brightness of the disc.

We show these contributions to the disc brightness in Figs. \ref{fig:sb_1um} and \ref{fig:sb_12um}, assuming CO-dominated secondary gas for the left panel ($\mu_{\rm g}=28$), and H$_2$-dominated, primordial gas for the right panel ($\mu_{\rm g}=2.3$). For the secondary gas, the gas masses (see plot legends) correspond to a range of estimated CO masses in gaseous debris discs, as compiled by \citet{Marino2020}. For the primordial gas, we assume that the gas masses are $10^3 \times$ higher, roughly corresponding to an H$_2$-CO number density ratio of $\sim 10^4$, as often assumed for protoplanetary discs. We find that the brightness of the unbound grains in the halo is significantly enhanced only for primordial gas, and even then only if the gaseous disc is fairly massive ($M_{\rm g}=100$\,M$_\oplus$). The gas surface density at a radius of 100\,au corresponding to the maximum considered gas masses is $\approx 10^{-4}$\,g\,cm$^{-2}$ for the CO-dominated gas and $\approx 10^{-1}$\,g\,cm$^{-2}$ for the H$_2$-dominated gas. The dimensionless subsonic stopping times for the blowout-sized grains are $T_{\rm ss} \sim$ 24 and 0.024, respectively.

The subsonic stopping times do not depend on the composition of the gas (i.e. the mean molecular weight) for equal gas masses. The mean molecular weight enters the calculations through the mean thermal velocity (or the sound speed). For CO-dominated gas, the mean thermal velocity is around 3.5 times lower, and consequently the midplane density is 3.5 times higher for the same surface density. The two effects cancel out in the expression for the subsonic stopping time. Since particles do reach supersonic velocities, the mean molecular weight is still somewhat relevant. The one remaining effect of a different mean molecular weight is (for unbound grains) a different radial velocity in Eq. (\ref{eq:v_r_term}), which affects the particle density at a given radius. Nevertheless, the very different gas masses expected for the CO-dominated and H$_2$-dominated gas are significantly more important than the gas composition.

The slope of the contribution of unbound grains to the surface brightness profile steepens in a gas-rich disc compared to that in a gas-poor disc. For scattered light, in the absence of gas or in a gas-poor disc, the surface brightness at large radii depends on the radius as $\propto r^{-3}$. This is because at large radii all particles reach their terminal velocity from Eq. (\ref{eq:v_r_nogas_inf}), making $v_r$ independent of radius. Then, in Eq. (\ref{eq:sb_unbound}), the only dependence of surface brightness on radius is the pre-factor $r^{-3}$, which comes from geometrical dilution of stellar flux hitting the grains ($r^{-2}$) and from geometrical dilution of the density of the radially escaping dust grains ($r^{-1}$). For the gas-rich, H$_2$-dominated gas disc with $M_{\rm g}=100$\,M$_\oplus$, the radial dependence for $r<500$\,au is better fit by $\propto r^{-5}$, if a power law is assumed. This steepness is explained by the exponential taper of the self-similar gas surface density profile that we assume. In this high-gas-mass limit, dust grains move through the gaseous disc at a subsonic velocity given by Eq. (\ref{eq:v_r_with_gas}), and the surface brightness of unbound grains depends on the radius approximately as $\propto r^{-3.5} \textrm{exp}(-r/r_{\rm belt})$ for the gas profile we adopted.

\section{Discussion and conclusions} \label{sec:conclusions}
The maximum gas mass considered in most of our calculations is 100\,M$_\oplus$, measured from the centre of the planetesimal belt out to large distances. Extrapolating the assumed surface density radial profile inwards to 0.1 \, au yields a total mass in the gaseous disc of 0.05\% of the assumed stellar mass, and this mass roughly corresponds to what might be the expected mass of a protoplanetary disc at late stages of its evolution. At such a gas mass, particles with $\beta \gtrsim 0.6$ remain unbound despite the gas drag. Even if the gas mass is increased to 1000 \,$M_\oplus$, the particles with $\beta \gtrsim 0.65$ remain unbound. However, the escape of the unbound grains is slowed down and their density is enhanced compared to a low-gas-mass disc. This is true not only within the gaseous disc but also after particles leave the disc, as previously found by \citet{Pearce2020}. For the latter effect, we find that the relevant quantity is the location of the outer gas edge, defined as the grain size-dependent radius at which the gas density drops sufficiently for the grain to decouple from the gas, that is, for the drag stopping time to rise to $T_{\rm s}=1$ (see Eq. \ref{eq:v_r_with_gas_infty}); the wider the gaseous disc, the lower the grain velocity at infinity. For lower gas masses, such as those expected if the observed CO in debris discs is of secondary origin, gas does not significantly affect the unbound grains and their contribution to the surface brightness of the disc.

Our results imply that, if the gas in cold debris discs is protoplanetary gas that has not yet been cleared out, gaseous debris discs should be more abundant in sub-blowout dust grains than gas-free discs. Some gaseous debris discs indeed appear to be overabundant in sub-micron dust, which has led to suggestions that gas drag slows the unbound grains \citep[e.g.][]{Bhowmik2019, Moor2019, Milli2023, Pawellek2024} and partly motivated our work. However, in some gaseous debris discs, the minimum grain size of the dust instead appears to be larger than the predicted blowout size \citep{Pawellek2019, Olofsson2023}. It is also not clear that an overabundance of sub-blowout grains preferentially occurs in gaseous discs. For example, a study of the colours of debris discs at short wavelengths found that discs generally tend to be blue, which is a signature of sub-micron grains being abundant \citep{Ren2023}. Comparing the discs in their sample with the known gas-bearing discs listed in Table A.1 of \citet{BergezCasalou2024}, we find that eight of the 23 discs in their sample have CO gas detections, yet overall these discs do not have bluer colours than the other discs in the sample. However, this comparison does not take into account the dependence on stellar type. CO gas is typically detected around A-type stars, around which radiation pressure is also stronger, meaning A-type stars should naturally contain fewer sub-micron grains compared to other stellar types. Accordingly, \citet{Ren2023} find that the colours of the discs of A-type stars are less blue than those of less luminous stars. Nevertheless, another recent study fitted the minimum dust grain size (and other disc parameters) to observations of 133 discs, and found that the sub-sample of gaseous discs does not differ from the wider sample \citep{Marshall2026}. 
Based on these findings, we can tentatively conclude that there is currently no evidence that gaseous discs preferentially retain their unbound grains, and therefore no evidence that observed gas-rich debris discs host vast amounts of protoplanetary gas. Some other mechanism may be responsible for the overabundance of sub-micron dust in general, such as high dust production in bright, highly collisional discs \citep{Thebault2019}.

One caveat to this conclusion is that some of the unbound grains may become collisional if their number density increases sufficiently. In the collisional models of bright debris discs \citep{Thebault2019}, the density of dust grains drops by a factor of only 5--30 at the blowout size. We find that in a gas-rich disc, the radial velocity of unbound particles can be slowed down by a similar or a higher factor. For dust grains that leave the disc, the density is inversely proportional to the radial velocity, and so it would increase by the same factor. This implies that their density might increase above the density of marginally bound grains. However, at that point, the lifetime and the density of the unbound grains would be limited by collisions with other unbound and bound particles, rather than by their escape due to radiation pressure, as assumed in our work. Collisions could further limit the effect of the gas on the contribution of unbound grains to the disc's surface brightness, even in a gas-rich disc.

Making firmer predictions for debris disc observations naturally requires modelling the dynamics of the much more abundant bound grains under the influence of gas, as well as their collisional evolution \citep[e.g.][]{Takeuchi2001,Thebault2005,Krivov2009,Olofsson2022}. The analytical model for estimating the number density of unbound grains presented in this work may help to incorporate unbound grains into such models in a computationally efficient way. In a kinetic collisional model of gas-free discs, \citet{vanLieshout2014} accounted for the contribution of unbound grains using the analytical expression in Eq. (\ref{eq:v_r_nogas}) and Eq. (\ref{eq:num_dens}) to calculate their radial velocity and number density as function of radius, respectively. This avoids the need for slow numerical integration of unbound grains' orbits. Analogously, in gaseous discs, we find that at any given radius the radial velocity of unbound grains is well approximated by combining Eq. (\ref{eq:v_r_nogas}) with Eq. (\ref{eq:v_r_term}) derived by \citet{Takeuchi2001}, and with Eq. (\ref{eq:v_r_with_gas_infty}). This combination of analytical expressions agrees well with results from numerical integration and should thus be helpful in the development of future comprehensive dynamical and collisional models of bound and unbound dust grains in gaseous debris discs.

\begin{acknowledgements}
MRJ acknowledges funding provided by the Institute of Physics Belgrade, through the grant by the Ministry of Science, Technological Development, and Innovations of the Republic of Serbia. TDP was supported by a UKRI Stephen Hawking Fellowship and a Warwick Prize Fellowship, the latter made possible by a generous philanthropic donation. MCW is supported by the Science and Technology Facilities Council grant UKRI 1198.
\end{acknowledgements}


\bibliographystyle{aa} 
\bibliography{bibliography} 

\end{document}